# A Vortex Dynamic Equation for Sediment Transport.

Ngatcha Ndengna Arno Roland*[1]

[1]*National Higher Polytechnic school of Douala, university of Douala, BP 2107, Douala, Cameroun.*

*Corresponding author(\*): arnongatcha@gmail.com.*

## Abstract

*This work presents a new vortex dynamic equation for quasi-geostrophic flows over strongly variable sediment bottoms. The equation considers erosion/deposition exchanges near the bottom and the geometrical changes of the bed interface, extending some of the most commonly used vortex dynamic equations in the literature. This new equation shows promise for further coastal engineering applications. The derivation of the vortex equation is presented in detail, including all physical and hydrodynamic mechanisms involved. Various variants of the vortex dynamic equation are also presented, which are particularly useful for multiphase turbulent flows. The advantages of our model are explained, and several observations are made.*

## 1. Introduction

The vortex dynamic for ideal incompressible fluid, has been introduced in 1858 by Helmholtz's [1] and extended in 1869 in theory of atoms by Thomsom (Lord Kelvin) [2]. To model sediment dynamics in multiphase turbulent flows, it is crucial to consider the relevant mechanisms for turbulence evolution, such as vortices. The existence of a universal 3D structure of vortices in quasi-geostrophic flows remains an open problem. In shallow environments, the interaction between vortices of different sizes is an essential mechanism for the evolution of turbulent coastal flows. The question of the creation of a vortex remains open and relevant for certain fluid dynamic applications [3], [4], [5], [6], [7]. In sediment transport, the question of vortex-induced sediment suspension is a topic of debate. Specifically, how is the vortex created over an erodible bottom, and which physical mechanisms are involved? Additionally, what role do the bed interface and the erosion/deposition exchanges play in the creation of the vortex? Several papers address the creation and destruction of vortices by considering the mechanism created by the variation in velocity [7], [5] either by the friction and sediment exchanges [4], either by only the bed interface geometrical changes and diffusion and friction terms (see the model developed by ). When the sediment at the bottom moves, sediment exchanges occur, inducing sediment suspension in the suspended layer. The vortex equations developed above, and even more vortex equations, are not applicable to address the creation or destruction of a vortex over a mobile bottom (see Fig: 3 in appendix). It is important to consider both the geometrical changes of the bed interface and sediment flux at the bottom in vortex dynamics. Additionally, we can also consider the diffusion term and friction source term. This work demonstrates how relevant mechanisms can be utilized to design an original vortex dynamic equation that addresses the dynamics of sediment in multiphase turbulent flows. The account of vortex structures in turbulent flow is crucial in sediment transport. The novelties presented here include: (i) the method used to derive the vortex equation and (ii) the significant number of physical and hydrodynamic mechanisms involved in the equation. We also present some variants derived from the vortex modeling. These variants are particularly interesting and promising.

This work examines the coupling between turbulent flow and dynamic vortices, taking into account the denser nature of the flow that arises the thermodynamic effects. Vortices play a crucial role in the spectral transfer of energy and enstrophy between scales in geophysical fluids. We present results on the modeling of the structure and dynamics of vortices in stratified flows, in both two and three dimensions. This work describes turbulent flow phenomena based on the concept of potential vorticity dynamics and the evolution of enstrophy, which is vortical energy across scales. Potential vorticity is commonly used in the study of mixing flows or multiphase flows [4], [6]. It is interesting to derive a potential vorticity equation in terms of sediment concentration and bed interface, as it can demonstrate the role played by sediment exchange effects and the geometric changes of the bed interface in the creation of the vortex.

This paper presents a mathematical model for vortex dynamics over mobile bottoms based on multiphase flows. The model serves as a unified basis for subsequent developments and can describe the motion of water due to sediment exchange and bed interface. Additionally, it can accurately describe geophysical multiphase turbulent flows combined with sediment transport. We have developed a new potential vorticity formula in terms of sediment flux, as sediment transport exchanges contribute to the creation of the vortex. The purpose of this work is to enhance geophysical flow modeling and address various sediment transport issues. The paper is structured as follows. Section 2 is devoted to presenting the basis equations and physical principles of mechanisms of creation/destruction of the vortex. In section 3, we derive a new vortex equation that account not only both geometrical changes of the bed interface and sediment flux exchanges, but also the effects of friction and diffusion. Some variants of the new equation are exposed in details and their advantages are given. We conclude this work and we give several remarks in section 4.

## 2. A general vortex dynamic equation: Basis equations and physical principles

To derive the vorticity equation for sediment/water mixing flow, we consider the momentum conservation equation, the density conservation equation of mixing NS equations

$$\frac{\partial \rho u_i}{\partial t} + \frac{\partial (\rho u_i u_j)}{\partial x_j} + \frac{\partial p}{\partial x_i} = \mathcal{F}_i, \quad \frac{\partial \rho}{\partial t} + \frac{\partial \rho u_i}{\partial x_i} = 0, \tag{1}$$

where is the source term that reads:

$$\mathcal{F} = \nabla . \tau + \mathbf{F} + f_\sigma, \tag{2}$$

where $\mathbf{F} = -\frac{K_{a,b}}{h^b} \rho |U|^a U$, $f_\sigma = \sigma \kappa \hat{n} \gamma_s$, $\tau = \frac{1}{2}\mu(\nabla U + \nabla U^t)$

The first RHS term in (2) is the viscous term (due to the denser nature of the flow), the second is the friction source term (expressed in turbulence context) and the last is surface tension force that acts on the bed interface. The last effect plays important role in the generation/destruction vortex as proved below.

In the surface tension force term, $\sigma$ is the surface tension coefficient, $\hat{n}$ is the interface unit normal vector and interface curvature is given by: $\kappa = -\nabla \cdot \left(\frac{\nabla c}{|\nabla c|}\right)$. The function $\gamma_s$ is given by: $\gamma_s = |\nabla c|$.

The 3D vorticity is defined as curl of the flow velocity vector $\boldsymbol{\omega} = curl(U)$, with $U = (u_i), i = 1, 2, 3$.

We use the following identities:

$$\begin{cases} \boldsymbol{\omega} = \nabla \times U, \\ (U \cdot \nabla)U = \nabla\left(\frac{1}{2}U \cdot U\right) - U \times \boldsymbol{\omega}, \\ \nabla \times (U \times \boldsymbol{\omega}) = -\boldsymbol{\omega}(\nabla \cdot U) + (\boldsymbol{\omega} \cdot \nabla)U - (U \cdot \nabla)\boldsymbol{\omega}, \\ \nabla \cdot \boldsymbol{\omega} = 0, \\ \nabla \times \nabla \phi = 0, \end{cases} \quad (3)$$

where $\phi$ is any scalar field.

The vorticity may be created or destroyed cannot annihilate the physical effects from the horizontal of vorticity. In complex turbulent flows, coherent structures may be turbulent (or in this case velocity) components are correlated with each in both space and time. We derived above an evolution equation for the variable measuring the horizontal velocity profile distortion via the averaged kinetic energy equation. The turbulence variable is naturally related to the flow vorticity. The periodic vortex structure is formed on the sediment bottom (ripple, sand wave, etc.) under the action of oscillatory, which affects the sediment motion.

For three-dimensional flow, the vorticity $\boldsymbol{\omega}[1/s]$ reads

$\boldsymbol{\omega} = curl\, U = (\omega_1, \omega_2, \omega_3) = \left((\varepsilon^2 \partial_y w - \partial_x v), (-\partial_z u + \varepsilon^2 \partial_x w), (\partial_x v - \partial_y u)\right)$

The vertical component of the vorticity describing the lateral spreading and shearing of the flow is given as follows:

$\omega_3 = \partial_x v - \partial_y u$.

While vorticity gives the general internal rotation of the water near the bed due the bedforms, the vertical component of the vorticity $\omega_2$ provides a more intuitive definition of vorticity. Its magnitude can be associated to the spinning of a paddle wheel that lies on the horizontal plane of the flow. Note that it is not that acts in the vorticity in shallow water environment. Often $\omega_2$ is smaller than the cross-stream vorticity component related to the no-slip condition that is $\omega_2 = -(\varepsilon^2 \partial_x w - \partial_z u) \approx \partial_z u$. From a physical point of view, $\partial_z u$ represents the rotation in the plane of the flow. The vortex can be positive (see Fig 1) or negative (see Fig 2). For antisymmetric flow, the situation where the vortex is negative is observed in Fig 1 and positive in Fig 2.

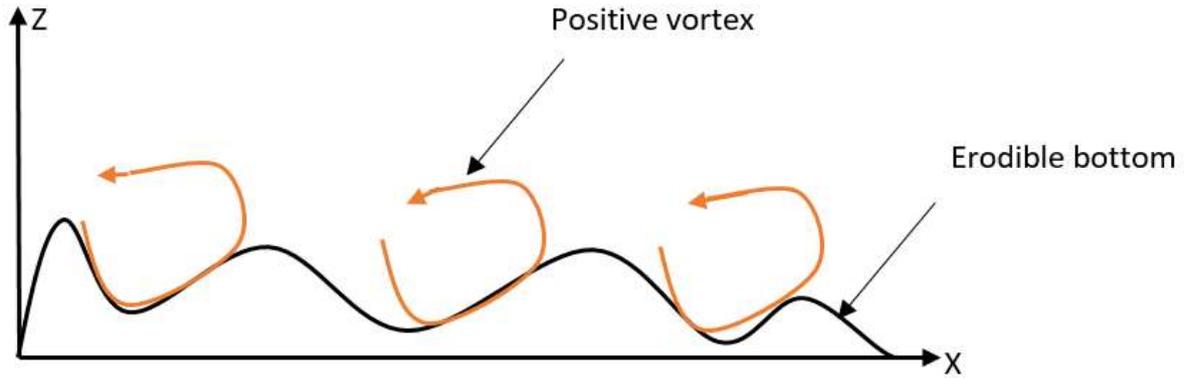

**Fig: 1** Positive vortex over erodible strongly variable bottom.

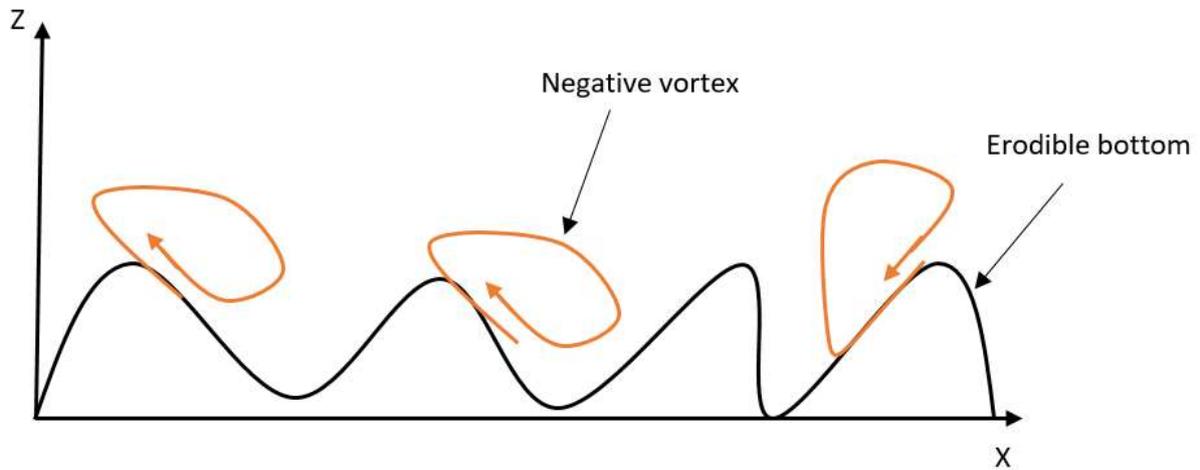

**Fig: 2** Negative vortex over strongly variable erodible bed.

From the mass conservation in Eq. (1), we obtain:

$$\frac{\partial u_j}{\partial x_j} = -\frac{1}{\rho}\frac{D\rho}{Dt}.$$

Accounting the vorticity, the momentum equation given in Eq. (1) becomes:

$$\frac{\partial U}{\partial t} + \omega \times U + \nabla\left(\frac{|U|^2}{2}\right) = \frac{1}{\rho}\mathcal{F} - \frac{1}{\rho}\nabla p, \qquad (4)$$

Taking *curl* of (4), according to the above identities (3), one obtain:

$$\frac{\partial \boldsymbol{\omega}}{\partial t} + \nabla \times (\boldsymbol{\omega} \times U) + \nabla \times \left\{ \nabla \left( \frac{|U|^2}{2} \right) \right\} = \nabla \times \left( \frac{1}{\rho} \mathcal{F} \right) - \nabla \times \left\{ \frac{1}{\rho} \nabla p \right\}, \quad (5)$$

The equation (5) shows that the vorticity can also increased akin to the dynamo effect by the $\nabla \times (\boldsymbol{\omega} \times U)$ term. This situation is analogous to the induction term in dynamo theory where $\boldsymbol{\omega}$ plays the role of the magnetic field. As observed in the previous relation the term $\nabla \times \left\{ \frac{1}{\rho} \nabla p \right\}$ emerges when taking the *curl* of the pressure gradient term $\nabla p$. Moreover it can proportional to the cross product of the gradients of pressure and density. Under certain conditions this term does not disappear. Using the fact that:

$$\nabla \times (A \times B) = A(\nabla . B) - B(\nabla . A) + A \times (\nabla \times B) + (B.\nabla)A - (A.\nabla)B \quad (6)$$

We have following (6)

$$\nabla \times (\boldsymbol{\omega} \times U) = \boldsymbol{\omega}(\nabla . U) + (U.\nabla)\boldsymbol{\omega} - (\boldsymbol{\omega} . \nabla)U,$$

the equation (5) becomes:

$$\frac{D\boldsymbol{\omega}}{Dt} = \boldsymbol{\omega} . \nabla U - \boldsymbol{\omega} \nabla . U - \nabla \times \left\{ \nabla \left( \frac{|U|^2}{2} \right) \right\} + \nabla \times \left( \frac{1}{\rho} \mathcal{F} \right) + \frac{1}{\rho^2} \nabla \rho \times \nabla p, \quad (7)$$

The term $\boldsymbol{\omega} \nabla . U$ proves that the convergence (divergence) of the mixing flow i.e. $\nabla . U$ will increases (decreases) its vorticity.

Let $\nabla \times \left( \frac{1}{\rho} \mathcal{F} \right) = \nabla \times \left( \frac{1}{\rho} \nabla . \tau \right) + \nabla \times \left( \frac{1}{\rho} \mathbf{F} \right) + \nabla \times \left( \frac{\sigma}{\rho} \kappa \hat{n} \gamma_s \right)$.

Considering the following equality,

$$\nabla(A.B) = (A.\nabla)B + (B.\nabla)A + A \times (\nabla \times B) + B \times (\nabla \times A). \quad (8)$$

We obtain:

$$\frac{D\boldsymbol{\omega}}{Dt} = (\boldsymbol{\omega} . \nabla)U - \boldsymbol{\omega} \nabla . U + \nabla^2 \boldsymbol{\omega} + \frac{1}{\rho^2}(\nabla \rho \times \nabla p) + \frac{1}{\rho^2} \frac{\rho_\infty}{Fr}(\nabla \rho \times \mathbf{g}) - \frac{1}{\rho^2} \nabla \rho \times \nabla . \tau \\ + \nabla \times \left( \frac{1}{\rho} \mathbf{F} \right) + \frac{\sigma}{\rho} \nabla \kappa \times \nabla c \quad (9)$$

Here, $\frac{D\boldsymbol{\omega}}{Dt}$ on the LHS represents the material derivative of $curl U$. It describes the rate of change of vorticity of the moving fluid particle. The term $\boldsymbol{\omega} . \nabla U$ represents the vortex-stretching or vortex contraction. The stretching vortex amplifies the vorticity when the velocity is diverging in the direction parallel to $\boldsymbol{\omega}$. It represents the contribution of shearing of velocity to the rate of change of vorticity. The term $\boldsymbol{\omega} \nabla . U$ is the expansion term, $\nabla^2 \boldsymbol{\omega}$ is the vortex diffusion, $\frac{1}{\rho^2}(\nabla\rho\times\nabla p)$ is the *pressure-torque*. This term correspond to the vorticity generation/destruction due to misalignment between the density gradient vector and pressure gradient.

The term $\frac{1}{\rho^2}\frac{\rho_\infty}{Fr}(\nabla\rho\times\mathbf{g})$ is the gravity term and $\nabla\times\left(\frac{1}{\rho}\nabla.\tau\right)$ is the viscous term that correspond to the vorticity generation/destruction due to misalignment between the density gradient vector and viscous stresses. It also named viscous torque. The term $\nabla\times\left(\frac{1}{\rho}\mathbf{F}\right)$ is the sum of external body forces. The last four terms of RHS of appear only due to nonhomogeneous-ness nature of the flow (or multiphase nature of the flow).

The term on the left represents the vortex convection. Because the flow is incompressible, the expansion terms cancel out, i.e. $\omega\nabla.U = O(\Delta)$. The turbulent kinetic energy cascades caused by low Reynolds very often occur in small regions close to the region where the kinetic energy dissipation rate is. It is therefore useful to consider minor vortex-stretch effects i.e. $\omega.\nabla U = O(\Delta)$. The stretching vortex amplifies the vorticity when the velocity is diverging in the direction parallel to $\omega$. The above vortex dynamic formulations with stretching vortex writes:

$$\frac{D\omega}{Dt} = (\omega.\nabla)U + \frac{1}{\rho^2}(\nabla\rho\times\nabla p) + \frac{1}{\rho^2}\frac{\rho_\infty}{Fr}(\nabla\rho\times\mathbf{g}) - \frac{1}{\rho^2}\nabla\rho\times\nabla.\tau + \nabla\times\left(\frac{1}{\rho}\mathbf{F}\right) + \frac{\sigma}{\rho}\nabla\kappa\times\nabla c. \quad (10)$$

The dynamic equilibrium of the sediment is therefore driven by the influence of viscosity gradients (viscous term), variable density gradients (baroclinic couples) and the effects of gravity (or stratification effect). In the diluted regions, the effect of gravity can be neglected i.e., $\frac{1}{\rho^2}\frac{\rho_\infty}{Fr}(\nabla\rho\times\mathbf{g})\approx 0(\Delta)$ and in this case the dynamic equation becomes:

$$\frac{D\omega}{Dt} = \frac{1}{\rho^2}(\nabla\rho\times\nabla p) - \frac{1}{\rho^2}\nabla\rho\times\nabla.\tau + \nabla\times\left(\frac{1}{\rho}\mathbf{F}\right) + \frac{\sigma}{\rho}\nabla\kappa\times\nabla c, \frac{D(.)}{Dt} = \frac{\partial(.)}{\partial t} + (U.\nabla)(.) \quad (11)$$

Here, the stratification effects are not really visible it can only indirectly contribute to vorticity production because a constant gravitational gives up when taking the *curl*.

When the water becomes hyperconcentrated, and, in presence of non-isothermal and non-entropy conditions, the gravity reduces, becomes non-constant and depends on density ($\mathbf{g'} = f(\rho)$ where $\mathbf{g'}$ is the reduced density). In such situation, the stratification effect does not vanish and $\frac{1}{\rho^2}\frac{\rho_\infty}{Fr}(\nabla\rho\times\mathbf{g}) \to \frac{1}{\rho^2}\frac{\rho_\infty}{Fr}(\nabla\rho\times\mathbf{g'})$.

Then the vorticity equation (11) take the form following:

$$\frac{D\boldsymbol{\omega}}{Dt} = \frac{1}{\rho^2}(\nabla\rho\times\nabla p) - \frac{1}{\rho^2}\nabla\rho\times\nabla.\tau + \nabla\times\left(\frac{1}{\rho}\mathbf{F}\right) + \frac{\sigma}{\rho}\nabla\kappa\times\nabla c + \frac{1}{\rho^2}\frac{\rho_\infty}{Fr}(\nabla\rho\times\mathbf{g}'), \tag{12}$$

The vorticity equation can be coupled with the new density-stratified liquid model introduced by [8]:

$$\begin{cases} \dfrac{\partial h\overline{\mathbf{u}}}{\partial t} + \nabla.\left(h\overline{\mathbf{u}}\right) = M_1 + M_2, \\ \dfrac{\partial \overline{\rho}h\overline{\mathbf{u}}}{\partial t} + \nabla.\left(\overline{\rho}h\overline{\mathbf{u}}\otimes\overline{\mathbf{u}}\right) + \nabla\overline{p} + \nabla\overline{p}' = h\overline{\mathbf{F}} + \rho\mathbf{u}(\eta)M_1 + \rho\mathbf{u}(Z_b)M_2, \\ \dfrac{\partial h\overline{\rho}}{\partial t} + \nabla.\left(h\overline{\mathbf{u}}\overline{\rho}\right) = \nabla.\left(hv'\nabla\overline{\rho}\right) + \overline{\rho}(\eta)M_1 + \overline{\rho}(Z_b)M_2, \\ \dfrac{\partial h\mathbf{R}}{\partial t} + \nabla.\left(h\overline{\mathbf{u}}\mathbf{R}\right) = \nabla.\left(hv'\nabla\mathbf{R}\right) + \mathbf{R}(\eta)M_1 + \mathbf{R}(Z_b)M_2, \end{cases} \tag{13}$$

Here, $h$ is the water depth, $h\mathbf{u}$ the water discharge, $\rho$ is the density and $\mathbf{R}$ is the added mass, $\overline{p}$ is the mean pressure while $\overline{p}'$ is the fluctuating pressure. The rest of terms are explained in appendix A (below).

By considering (sufficiently smooth) domain $\Omega \subset \mathbb{R}^2$ and assume that

$h.\mathbf{n} = 0$, $h\overline{\rho}.\mathbf{n} = 0$, $\overline{\rho}h\overline{\mathbf{u}}.\mathbf{n} = 0$ $h\mathbf{R}.\mathbf{n} = 0$, on $\partial\Omega$, $\forall t \in ]0,T]$

where $\mathbf{n}: \partial\Omega \to \mathbb{R}^d$ denotes the mapping that assigns the outward unit normal vector to any $x \in \partial\Omega$.

This stratified model (13) allows one to know how it possible to describe the motion of a dense water in coastal environment. The model extends those based on diluted or weakly diluted flows and allows to give a real definition of a hyperconcentrated water flow rigorously. The derivation of the model is presented in appendix. The stratification effects become visible it can directly contribute to vorticity production because a constant gravitational gives up when taking the *curl*. When the condition of isothermality or adiabaticity are not relaxed we can neglect the baroclinic effect and the vorticity equation becomes:

$$\frac{D\boldsymbol{\omega}}{Dt} = -\frac{1}{\rho^2}\nabla\rho\times\nabla.\tau + \nabla\times\left(\frac{1}{\rho}\mathbf{F}\right) + \frac{\sigma}{\rho}\nabla\kappa\times\nabla c + \frac{1}{\rho^2}\frac{\rho_\infty}{Fr}(\nabla\rho\times\mathbf{g}') \quad , \quad \frac{D(.)}{Dt} = \frac{\partial(.)}{\partial t} + (U.\nabla)(.)$$
(14)

This situation states that in absence of *pressure-torque* effect, the gravity reduced effect, the geometrical changes of interface, the friction term and viscosity effect can lead to vorticity production. In order to estimate the production of vorticity, one could derive an evolution equation for the enstrophy density $\psi = \frac{1}{2}\boldsymbol{\omega}^2$ (vorticity squared) by the dot product of (9) with $\boldsymbol{\omega}$ and use a closure assumption for the resulting triple correlation. The enstrophy represents a scalar quantity that intrinsically reflects the strength of the vorticity field without a vector implication. The enstrophy can help to measure vortical energy of turbulent structures. According to equation the enstrophy density equation is given by:

$$\frac{D\psi}{Dt} = \omega.(\omega.\nabla U) - \omega.(\omega\nabla.U) + \omega.\nabla^2\omega + \omega.\left(\frac{1}{\rho^2}(\nabla\rho\times\nabla p)\right) + \omega.\left(\frac{1}{\rho^2}\frac{\rho_\infty}{Fr}(\nabla\rho\times\mathbf{g}')\right) - \omega.\left(\frac{1}{\rho^2}\nabla\rho\times\nabla.\tau\right)$$

$$+ \omega.\left(\nabla\times\left(\frac{1}{\rho}\mathbf{F}\right)\right) + \omega.\left(\frac{\sigma}{\rho}\nabla\kappa\times\nabla c\right)$$

(15)

In this equation, all the RHS terms represent the rate of enstrophy generation/destruction by the related mechanisms involved in vortex dynamic equation given by (9). The rate of enstrophy production by the surface tension balances the disruptive mechanism of vortex stretching.

**A multiphase turbulence model accounting morphodynamics**

We can describe the multiphase turbulence phenomena based on the concept of vorticity transport and the evolution of enstrophy. A three-dimensional hyperconcentrated flow model describing the multiphase turbulence phenomena combined with sediment transport can be given as follows:

$$\begin{cases} \frac{D\omega}{Dt} = (\omega.\nabla)U - \omega\nabla.U + \nabla^2\omega + \frac{1}{\rho^2}(\nabla\rho\times\nabla p) + \frac{1}{\rho^2}\frac{\rho_\infty}{Fr}(\nabla\rho\times\mathbf{g}') - \frac{1}{\rho^2}\nabla\rho\times\nabla.\tau \\
\qquad\qquad + \nabla\times\left(\frac{1}{\rho}\mathbf{F}\right) + \frac{\sigma}{\rho}\nabla\kappa\times\nabla c, \qquad\qquad in\ \Omega\times]0,T] \\
\frac{D\psi}{Dt} = \omega.(\omega.\nabla U) - \omega.(\omega\nabla.U) + \omega.\nabla^2\omega + \omega.\left(\frac{1}{\rho^2}(\nabla\rho\times\nabla p)\right) + \omega.\left(\frac{1}{\rho^2}\frac{\rho_\infty}{Fr}(\nabla\rho\times\mathbf{g}')\right) - \omega.\left(\frac{1}{\rho^2}\nabla\rho\times\nabla.\tau\right) \\
\qquad\qquad + \omega.\left(\nabla\times\left(\frac{1}{\rho}\mathbf{F}\right)\right) + \omega.\left(\frac{\sigma}{\rho}\nabla\kappa\times\nabla c\right), \quad in\ \Omega\times]0,T] \\
\omega(X,0) = \omega^0(X),\ \ \psi(X,0) = \psi(X), \qquad in\ \Omega \\
\omega.\mathbf{n} = 0, \qquad \psi.\mathbf{n} = 0 \qquad in\ \Omega\times]0,T]
\end{cases}$$

(16)

The vortex stretching/telting and the surface tension terms can be dominating in the compute of the rate enstrophy. A simple model can be given for homentropic weakly diluted flows without friction and viscosity effects. One has:

$$\begin{cases} \frac{D\omega}{Dt} = (\omega.\nabla)U + \frac{\sigma}{\rho}\nabla\kappa\times\nabla c, \qquad in\ \Omega\times]0,T] \\
\frac{D\psi}{Dt} = \omega.(\omega.\nabla U) + \omega.\left(\frac{\sigma}{\rho}\nabla\kappa\times\nabla c\right), \qquad in\ \Omega\times]0,T] \\
\omega(X,0) = \omega^0(X),\ \ \psi(X,0) = \psi(X), \qquad in\ \Omega \\
\omega.\mathbf{n} = 0, \qquad \psi.\mathbf{n} = 0. \qquad in\ \Omega\times]0,T]
\end{cases} \qquad (17)$$

The contribution of enstrophy production and destruction rates by vortex stretching and surface tension depends on the wavenumber. It has been demonstrated that $\boldsymbol{\omega} \cdot \left( \dfrac{\sigma}{\rho} \nabla \kappa \times \nabla c \right)$ is negative for low wavenumbers and becomes positive at higher ones. This model can be improved or simplified to represent more phenomena. In 2D flows, vortex stretching and tilling are absent, causing two-dimensional turbulence to behave differently from three-dimensional turbulence.

## 3. Potential Vorticity (PV) theorem for sediment transport dynamics. Development of a new kind PV accounting sediment flux and geometrical changes of bed interface.

The sediment exchanges near the bottom can modify the motion of the vorticity. Furthermore, the geometrical changes of bed interface (via the surface tension force) can contribute to the creation/destruction of the vorticity. The literature does not provide a complete description of the vortex in the theory of sediment transport. It is well known that fluid rotations near the bottom involve particle transfer. Furthermore, the moving bottom can cause water rotations due to its wave nature. It is important to develop a situation where the vorticity equation describes both sediment transport and morphodynamics. A potential vorticity (PV) equation is rigorously established to demonstrate the possibility.

This section establishes a PV equation that demonstrates the crucial role of sediment exchange and bottom movement in the generation and destruction of vorticity. Two scalar operators of the form are defined:

$$\Pi_\varphi = \frac{\boldsymbol{\omega}}{\rho} \nabla \varphi, \quad \widehat{\Pi}_\varphi = \frac{\boldsymbol{\omega}}{\rho} \frac{\nabla \varphi}{|\nabla \varphi|} = \frac{\boldsymbol{\omega}}{\rho} \hat{n}, \tag{18}$$

where $\varphi$ is an arbitrary scalar that can be the potential temperature, potential density of water/sediment, etc. Ertel's PV theorem [3] can provide a description the physics of a sediment/water mixing motion. Same Ertel's PV theorem has been used in several works (see [5] [6], [4]). There is an analogy between Ertel's theorem and Kelvin's theorem [5]. Ertel's theorem is a differential statement of Kelvin's theorem where the contour is chosen in a surface for which the *pressure-torque* $\dfrac{1}{\rho} \nabla \ln \rho \times \nabla p$ lies in the surface and does not contribute to the change in the circulation.

We assume that suspended sediment concentrated equation is considered as a scalar water property that is not materially conserved such that:

$$\frac{Dc}{Dt} = \Sigma, \tag{19}$$

where $\Sigma$ is a source/sink term for $c$, and thus is associated to sediment exchange processes near the bed or at the bottom interface $z = Z_b$. When any scalar is conserved i.e. $\Sigma = 0$ then homentropic flow conserves $\Pi_\varphi$. In presence of sediment exchange and interface geometrical changes the PV is not conserved.

According the mass conservation equation the divergence reads $\nabla . U = -\frac{1}{\rho}\frac{D\rho}{Dt} = -\frac{D\ln(\rho)}{Dt}$.

By considering the equality $\frac{D}{Dt}\left(\frac{\omega}{\rho}\right) = \frac{1}{\rho}\frac{D\omega}{Dt} - \frac{\omega}{\rho^2}\frac{D\rho}{Dt}$ and recognizing the above general vorticity equation (10) we obtain:

$$\frac{D}{Dt}\left(\frac{\omega}{\rho}\right) = (\frac{\omega}{\rho}.\nabla)U + \frac{1}{\rho}\nabla^2\omega + \frac{1}{\rho^2}(\nabla \ln \rho \times \nabla p) + \frac{1}{\rho^3}\frac{\rho_\infty}{Fr}(\nabla \rho \times \mathbf{g'}) - \frac{1}{\rho^2}\nabla \ln \rho \times \nabla . \tau \\ + \frac{1}{\rho}\nabla \times \left(\frac{1}{\rho}\mathbf{F}\right) + \frac{\sigma}{\rho^2}\nabla \kappa \times \nabla c \qquad (20)$$

$\frac{\omega}{\rho}$ is frozen into the fluid since it obeys the same equation as a field of infinitesimal displacement vectors between fluid particles. Furthermore, since the velocity field is continuous, the distortion $\frac{\omega}{\rho}$ is continuous. $\frac{\omega}{\rho}$ can never be torn apart. Its topology is preserved despite distortion.

To write the vorticity in terms of potential, some manipulations are required. Now we provide a special and temporal variation of the PV given by (18) in terms of mass sediment conservation and source term (diffusion and bed friction). To obtain this evolution equation, we multiply (20) by $\nabla c$ where $c = c(x,y,z,t)[Kg/m^3]$ is sediment concentration. One obtain:

$$\nabla c . \frac{D}{Dt}\left(\frac{\omega}{\rho}\right) = \nabla c . \left[\frac{1}{\rho}\left(\varepsilon^2 \partial_y w - \partial_x v\right)\partial_x U + \frac{1}{\rho}\left(-\varepsilon^2 \partial_z u + \partial_x w\right)\partial_y U + \frac{1}{\rho}\left(\partial_x v - \partial_y u\right)\partial_z U\right] + \\ + \nabla c . \left(\frac{1}{\rho}\nabla \times \left(\frac{1}{\rho}\mathcal{F}\right)\right) + \nabla c . \left(\frac{1}{\rho^2}\nabla \ln \rho \times \nabla p\right), \qquad (21)$$

We see well the influence of shearing of velocity to the rate of change of vorticity. After some tedious algebra equation (21) gives:

$$\frac{D}{Dt}\left(\frac{\omega}{\rho}\nabla c\right) = \frac{\omega}{\rho}\nabla \Sigma + \frac{\sigma}{\rho^2}\nabla c . (\nabla \kappa \times \nabla c) + \frac{\nabla c}{\rho}\left(\nabla \times \left(\frac{1}{\rho}\mathbf{F}\right)\right) - \frac{1}{\rho^2}\nabla c . (\nabla \ln \rho \times \nabla . \tau) + \frac{1}{\rho^2}\nabla c . (\nabla \ln \rho \times \nabla p), \quad (22)$$

When we consider only the friction term and erosion/deposition exchanges via $\nabla \Sigma$, we retrieve the formulation of the Ertel PV theorem, able to describe a sediment-laden river jet system as in [4]. The term $\frac{\nabla c}{|\nabla c|}$ is more relevant than $\nabla c$ for showing the bed interface movement. We can therefore reformulate the Ertel PV theorem to describe the geometrical changes of the bed interface and this is a particular novelty of this derivation.

On the normal of bed interface represented by $\hat{n} = \frac{\nabla c}{|\nabla c|}$, we can reformulate above relation (22). By dividing (22) by $|\nabla c|$, it becomes:

$$\frac{D}{Dt}\left(\frac{\omega}{\rho}\hat{n}\right)=\frac{\omega}{\rho}\nabla\left(\frac{\Sigma}{|\nabla c|}\right)+\frac{\sigma}{\rho^2}(\nabla\kappa\times\nabla c).\hat{n}+\frac{1}{\rho}\left(\left(\nabla\times\left(\frac{1}{\rho}\mathbf{F}\right)\right).\hat{n}\right)-\frac{1}{\rho^2}\hat{n}.(\nabla\ln\rho\times\nabla.\tau)+\frac{1}{\rho^2}\hat{n}.(\nabla\ln\rho\times\nabla p),$$
(23)

and this leads to:

$$\frac{D\widehat{\Pi}_c}{Dt}=\frac{\omega}{\rho}\nabla\left(\frac{\Sigma}{|\nabla c|}\right)+\frac{\sigma}{\rho^2}(\nabla\kappa\times\nabla c).\hat{n}+\frac{1}{\rho}\left(\left(\nabla\times\left(\frac{1}{\rho}\mathbf{F}\right)\right).\hat{n}\right)-\frac{1}{\rho^2}\hat{n}.(\nabla\ln\rho\times\nabla.\tau)+\frac{1}{\rho^2}\hat{n}.(\nabla\ln\rho\times\nabla p),$$
(24)

We can reasonably suppress the term $\frac{\nabla c}{|\nabla c|}.\nabla\ln\rho\times\nabla p$ under isothermally assumption. In this case equation (24) becomes:

$$\frac{D\widehat{\Pi}_c}{Dt}=\frac{\omega}{\rho}\nabla\left(\frac{\Sigma}{|\nabla c|}\right)+\frac{\sigma}{\rho^2}(\nabla\kappa\times\nabla c).\hat{n}+\frac{1}{\rho}\left(\left(\nabla\times\left(\frac{1}{\rho}\mathbf{F}\right)\right).\hat{n}\right)-\frac{\hat{n}}{\rho^2}.(\nabla\ln\rho\times\nabla.\tau) \quad (25)$$

The equation (25) represents the vortex dynamic modeling developed in this work and is particularly interesting for further investigations. This equation proves that the geometrical changes (due to movement of the bed interface) arise the vorticity near the bottom. To close the general vortex dynamic equation exposed in (25), we give an expression of the friction source term $\nabla\times\left(\frac{1}{\rho}\mathbf{F}\right)$.

Here we assume that the bottom friction acts only in horizontal direction. We begin by giving a more general friction term as:

$$\mathbf{F}=-K_{a,b}\frac{\rho}{h^b}|U|^a U, \quad a,b\in\mathbb{R}, \quad K_{a,b}\in\mathbb{R}_+^* \quad (26)$$

We have:

$$\left[\nabla\times\frac{\mathbf{F}}{\rho}\right]=\left(0,-K_{a,b}\frac{|U|^a}{h^b}\frac{\partial u}{\partial z},-K_{a,b}\frac{|U|^a}{h^b}\left(\frac{\partial v}{\partial x}-\frac{\partial u}{\partial y}\right)\right) \quad (27)$$

For, $a=b=1$, if we consider that $u\gg v$ and $\nabla u\gg\nabla v$, $\nabla v\ll\varepsilon$ then,

$$\left[\nabla\times\frac{\mathbf{F}}{\rho}\right]=\left(0,-2K\frac{u}{h}\frac{\partial u}{\partial z},-2K\frac{u}{h}\frac{\partial u}{\partial y}\right).$$

Then, $\frac{1}{\rho}\left[\nabla\times\frac{\mathbf{F}}{\rho}\right].\hat{n}=-2K\frac{u}{h\rho}\frac{1}{|\nabla c|}\frac{\partial u}{\partial z}\frac{\partial c}{\partial y}-2K\frac{u}{h\rho}\frac{1}{|\nabla c|}\frac{\partial u}{\partial y}\frac{\partial c}{\partial z}=-2K\frac{u}{h}\widehat{\Pi}_c.$

**Advantages of the vortex dynamic model (25)**

- The derived model given by (25) and (27) allows one to know how it possible to describe flows exhibiting vortex structures over erodible bed only via sediment exchanges, friction and geometrical changes of bed interface.
- The proposed vortex equation show the influence (or the effects) of the bed movement into the creation/destruction of the vortex.
- The vortex dynamic equation proposed here extends several vortex models developed in the literature and most used in sediment transport.
- The potential vorticity developed here is most relevant for representing the vortex over erodible bottom and its effect in the sediment transport.

**Other forms of the model**

If we neglect the friction and viscous terms, we have

$$\frac{D\widehat{\Pi}_c}{Dt} = \frac{\boldsymbol{\omega}}{\rho}\nabla\left(\frac{\Sigma}{|\nabla c|}\right) + \frac{\sigma}{\rho^2}\left((\nabla\kappa.\hat{n})\times\nabla c\right) \qquad (28)$$

This equation demonstrates that only sediment exchange and movement at the bed interface contribute to the generation and destruction of the vortex. This is a significant contribution to geophysical flow with a free surface. If we also assume no sediment exchange $\Sigma = 0$, our PVs becomes:

$$\frac{D\widehat{\Pi}_c}{Dt} = \frac{\sigma}{\rho^2}\left((\nabla\kappa.\hat{n})\times\nabla c\right) \qquad (29)$$

This formulation proves that the PVs is strongly influenced by the geometrical changes interface. This viewpoint is well confirmed by the rapid movement of the bottom during the motion of the water. Eq. (29).

Expressing the material derivative as a function of sediment exchange at the bottom interface and according to the kinematic equation above given in appendix (see Eq. (B4)), we can represent more generally $\Sigma$ in term of mass sediment exchange at the bottom interface in Eulerian configuration as:

$$\Sigma = D - E = \frac{Dc}{Dt} = (1-c)^n w_s + \varepsilon_s \frac{\partial c}{\partial z}, \qquad z = Z_b(x,y,t) \qquad (30)$$

which expresses the mass conservation term in Eq. (24) as the product of sediment concentration with fall velocity plus the (downward) settling of sediments. The potential vorticity equation given by (24) and (30) is valid only near the sediment bed. It disappears in free surface located at $z = \eta(x,y,t)$.

Considering the pressure gradient's dependence on density and its fluctuations, it is reasonable to assume that the pressure-torque term is not always zero. Therefore, the pressure-torque effect does not disappear, which is a unique characteristic of dense flow in long wave theory. Despite the water becoming dense, several published works neglect this effect. In this flow situation, collisional friction produces significant thermodynamic effects.

Equation (24) presents a general formulation based on Ertel's PV theorem that can represent sediment transport in turbulent flow with vortices affected by strong pressure gradients due to strong density (which results in added mass), external and internal forces due to bottom friction and viscous terms, as well as erosion/deposition exchanges. We can express the term $\nabla \Sigma$ in terms of sediment concentration as:

$$\nabla \Sigma = \nabla\left(w_s(1-c)^n\right) + \nabla\left(\varepsilon_s \frac{\partial c}{\partial z}\right) = w_s \nabla(1-c)^n + \nabla\left(\varepsilon_s \frac{\partial c}{\partial z}\right)$$
$$= \left( w_s \frac{\partial (1-c)^n}{\partial x} + \frac{\partial}{\partial x}\left(\varepsilon_s \frac{\partial c}{\partial z}\right), w_s \frac{\partial (1-c)^n}{\partial y} + \frac{\partial}{\partial y}\left(\varepsilon_s \frac{\partial c}{\partial z}\right), w_s \frac{\partial (1-c)^n}{\partial z} + \frac{\partial}{\partial z}\left(\varepsilon_s \frac{\partial c}{\partial z}\right) \right) \quad (31)$$

Here, the term $\nabla\left(\varepsilon_s \frac{\partial c}{\partial z}\right)$ is the gradient of the downward sediment flux. The above equation is most relevant for hyperconcentrated water flow.

## A simplified vortex dynamic model

Neglecting the vertical velocity as well as any vertical variation of the cross stream velocity i.e. $w, \partial_z v \approx 0$, $\boldsymbol{\omega} = curl U = (\omega_1, \omega_2, \omega_3) = \left(0, (\partial_z u + 0), (0 + \partial_y u)\right)$, we obtain a new expression of the potential vorticity $\Pi_c [s^{-1}/m]$.

$$\Pi_c = \boldsymbol{\omega} \frac{\nabla c}{\rho} = \frac{1}{\rho}\left(\frac{\partial u}{\partial z}\frac{\partial c}{\partial y} + \frac{\partial u}{\partial y}\frac{\partial c}{\partial z}\right). \quad (32)$$

Neglecting the horizontal variation of the sediment concentration due to vertical rotation of the vortex we can express the PV theorem as:

$$\Pi_c = \boldsymbol{\omega} \frac{\nabla c}{\rho} = \frac{1}{\rho}\left(\frac{\partial u}{\partial y}\frac{\partial c}{\partial z}\right)$$

It is also relevant to express the evolution of the new PV given by $\widehat{\Pi}_c$ in terms of only sediment concentration flux. Assuming that the flow is weakly dense, we start by writing:

$$\frac{1}{|\nabla c|}\nabla \Sigma = w_s \hat{n} + \frac{1}{|\nabla c|}\nabla\left(\varepsilon_s \frac{\partial c}{\partial z}\right)$$

Then, we can reformulate the vortex dynamic equation given by Eq.(28) as follows:

$$\frac{D\widehat{\Pi}_c}{Dt} = w_s \widehat{\Pi}_c + \frac{\boldsymbol{\omega}}{\rho}\nabla\left(\frac{1}{|\nabla c|}\left(\varepsilon_s \frac{\partial c}{\partial z}\right)\right) + \frac{\sigma}{\rho^2}\left((\nabla \kappa.\hat{n})\times \nabla c\right) - 2K\frac{u}{h}\widehat{\Pi}_c \quad (33)$$

After a simple manipulation and neglecting the term $-\frac{\varepsilon_s}{|\nabla c|}\frac{\partial c}{\partial z}\nabla\left(\frac{\boldsymbol{\omega}}{\rho}\right)$ and assuming that the Schmidt number is constant, we can the equation (33) reformulates as follows:

$$\frac{D\widehat{\Pi}_c}{Dt} = w_s \widehat{\Pi}_c + \varepsilon_s \nabla \widehat{\Pi}_c + \frac{\sigma}{\rho^2}(\nabla \kappa \times \nabla c).\hat{n} - 2K\frac{u}{h}\widehat{\Pi}_c \qquad (34)$$

This equation (34) represents a formulation of the PV theorem to describe a sediment transport.

The model is also able to describe a sediment-laden river jet system due to both lateral and bottom shear stresses as well as depositional erosive effect. Via this formulation, we can explore potential effect of collisional on sediment due to dense nature of the flow. The collisional effect increase the shear stress and this later is most important than deposition. It is seem that it is possible to compact the nonhomogeneous Navier-Stokes equations combined with sediment transport into one equation capable describing the vortex evolution arisen from the sediment exchanges, friction term and geometrical changes of the bed interface.

When we remove the effect of geometrical changes and potential vorticity gradient, we obtain:

$$\frac{D\widehat{\Pi}_c}{Dt} = \left(-2K\frac{u}{h}\right)\widehat{\Pi}_c. \qquad (35)$$

We retrieve a similar model derived in [4] that states our PV changes is governed by frictional effects only. With that, an exact solution can be found at steady states.

**General remarks: Vortex motion over erodible bed**

The nature of vortex dynamic flow is often unclear and subject to interpretation. However, there are some generally accepted aspects, including:

**(i)** The occurrence of strong vortex dynamics in free surface flows when there is an abrupt and moving topography that varies significantly from one point to another.

**(ii)** Turbulence over erodible bed is associated to vorticity. Vorticity may be caused by sediment exchange and geometrical changes of the bed interface.

**(iii)** The turbulent flow of water near the bottom can become hyperconcentrated, resulting in strong density gradients and pressure fluctuations and in this case, the pressure-torque in vortex cannot neglected due to thermodynamic effects that arise.

**(iv)** $\frac{\boldsymbol{\omega}}{\rho}$ is frozen and its topology is preserved. Furthermore, since the velocity field is continuous, $\frac{\boldsymbol{\omega}}{\rho}$ is continuous and remains tangent to the moving bed interface of the normal $\hat{n}$.

**(v)** Vortex motion over erodible bottoms often exhibit high levels of intermittency. A vortex motion is intermittent if its variability is dominated by infrequent large events.

## 4. Conclusion

This work proposes a new fluid dynamics equation to describe vortex-induced sediment suspension over an erodible bottom. The motion of dense water is considered, and when the flow becomes dense, strong sediment-sediment interactions can cause thermodynamic effects. In such situations, pressure-torque can become important and may influence the dynamic vortex near the bottom. The analysis performed here shows that a turbulent flow over an erodible bed is associated with vorticity. Vorticity is a relevant concept in fluid mechanics, especially in sediment transport. It has been concluded that near the bottom, sediment exchanges and geometrical changes create vorticity. Additionally, friction and diffusion effects play a significant role in the creation and destruction of the vortex. This work proposes a new dynamic equation for vortices in sediment transport theory, specifically for situations where water flow is dense and accelerated. This text presents various versions of a dynamic vortex model and provides detailed explanations of the advantages of each variant. The equation is unique because it considers both sediment exchange and geometrical changes of the bed interface, which is a novelty in geophysical turbulent flow modeling with sediment transport. The proposed work improves upon several relevant studies in fluid mechanics and addresses many of the modeling insufficiencies found in the literature.

## Appendix A: The new density-stratified liquid model

We consider the system of equation following:

$$\begin{cases} \dfrac{\partial \rho}{\partial t} + \dfrac{\partial (\rho u_i)}{\partial x_i} = 0, \\ \dfrac{\partial \rho u_i}{\partial t} + \dfrac{\partial (\rho u_i u_j)}{\partial x_j} + \dfrac{\partial p}{\partial x_i} = \mathcal{F}_i, \\ \dfrac{\partial u_i}{\partial x_i} = 0, \end{cases} \quad (A1)$$

where $u_i$, $i = 1, 2, 3$ or $U = u_i = (u, v, w) = (\mathbf{u}, w)$ are the 3D mixing velocity components, $\rho$ is the mixing density and $p$ is the pressure term.

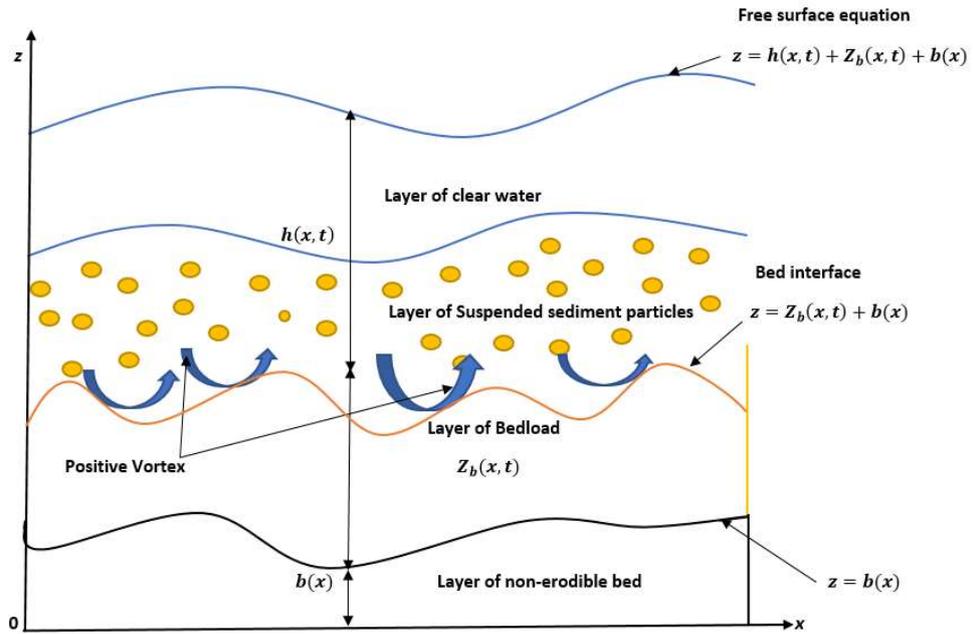

**Fig 3:** Zones in the flow modeling and parameters of the problem.

## Kinematic and boundary equations

A position of a fluid particle in a domain $\Omega \subset \mathbb{R}$ (see Fig 3) at the time $t$ is defined as follows:

$$\mathcal{X} = \left\{ (x, y, z), (x, y) \in \mathbb{R}^2, \ Z_b(x, y, t) \leq z \leq \eta(x, y, t), \ t \in \mathbb{R}_+^* \right\}.$$

where $\eta$ is a smooth function.

A point on the free surface writes:

$$M(x,y,z,t) = -z + \eta(x,y,t),$$

Assuming that any particle that is on the free surface at the initial instant will remain so at all instants, we have:

$$\frac{DM}{Dt} = 0, \qquad \frac{D(.)}{Dt} = \frac{\partial(.)}{\partial t} + (\mathbf{u}.\nabla)(.).$$

At the bottom surface or bed interface, one has $z = Z_b(x,y,t)$. Therefore, we can define the kinematic boundary conditions on both moving surfaces.

On the free surface we have:

$$\frac{\partial \eta}{\partial t} + (\mathbf{u}.\nabla)\eta = M_1 + u_3(\eta), \tag{A2}$$

where $M_1$ plays the role of entrainment processes near the free surface and is responsible for the mixing between layers.

On the bed interface we have:

$$\frac{\partial Z_b}{\partial t} + \mathbf{u}(Z_b)\nabla Z_b = M_2 + u_3(Z_b), \tag{A3}$$

where term $M_2$ can represent the exchange processes near the bottom.

The Leibniz's relations are also used to derive the new density-stratified liquid model:

$$\frac{\partial h\overline{\psi}}{\partial \mathbf{x}} = \frac{\partial}{\partial \mathbf{x}} \int_{Z_b}^{\eta} \psi \, dz = \int_{Z_b}^{\eta} \frac{\partial \psi}{\partial \mathbf{x}} dz - \psi(\eta)\frac{\partial \eta}{\partial \mathbf{x}} + \psi(Z_b)\frac{\partial Z_b}{\partial \mathbf{x}}, \quad \mathbf{x} = (x,y) \tag{A4}$$

and

$$\frac{\partial h\overline{\psi}}{\partial t} = \frac{\partial}{\partial x} \int_{Z_b}^{\eta} \psi \, dz = \int_{Z_b}^{\eta} \frac{\partial \psi}{\partial t} dz - \psi(\eta)\frac{\partial \eta}{\partial x} + \psi(Z_b)\frac{\partial Z_b}{\partial t}. \tag{A5}$$

Let us integrate the nonhomogeneous Navier-Stokes equations with respect to $z$ from $Z_b$ to $h$, we obtain:

$$\begin{cases} \overline{\dfrac{\partial \rho}{\partial t} + \dfrac{\partial (\rho u_i)}{\partial x_i}} = 0, \\ \overline{\dfrac{\partial \rho u_i}{\partial t} + \dfrac{\partial (\rho u_i u_j)}{\partial x_j} + \dfrac{\partial p}{\partial x_i}} = \overline{\mathcal{F}_i}, \\ \overline{\dfrac{\partial u_i}{\partial x_i}} = 0, \end{cases} \tag{A6}$$

To obtain the generalized shallow water-based equations, we apply an average along the depth of each equation in the system and use Leibniz's formula to simplify them into 2D equations.

The averaging free divergence equation gives:

$$\frac{\partial h\overline{\mathbf{u}}}{\partial t} + \nabla \cdot \left(h\overline{\mathbf{u}}\right) = M_1 + M_2 + ..., \qquad (A7)$$

The new momentum conservation law equations writes:

$$\frac{\partial \overline{\rho} h \overline{\mathbf{u}}}{\partial t} + \nabla \cdot \left(\overline{\rho} h \overline{\mathbf{u}} \otimes \overline{\mathbf{u}}\right) + \nabla \overline{p} + \nabla \overline{p'} = h\overline{\mathbf{F}} + \rho \mathbf{u}(\eta) M_1 + \rho \mathbf{u}(Z_b) M_2 + ..., \qquad (A8)$$

where $\overline{p} = \frac{gh^2 \overline{\rho}}{2} > 0$, $\overline{p'} = \frac{gh^2}{2}\left(\frac{\mathbf{R}}{\overline{\rho}}\right) > 0$, $\mathbf{R} = \overline{\rho' \rho'} > 0$

$\mathbf{R}[Kg^2/m^6]$ is the order of $\rho^2$, i.e. $O(\mathbf{R}) = O(\rho^{*2})$.

Here, $\mathbf{m} \otimes \mathbf{n}$ denotes the second order tensor with the components $(\mathbf{m} \otimes \mathbf{n})_{ij} = m_i n_j$.

We remark that the account both mean pressure and averaged fluctuating pressure in the model and this is particularly interesting. Both pressures are positive when $h > 0$, $\rho > 0$, $\mathbf{R} > 0$. Numerically the positivity of the water depth and the mixing density as well as the added mass can be ensured. For classical stratified liquid model as those developed by [7], [9] the term $\overline{p'} = \frac{gh^2}{2}\left(\frac{\mathbf{R}}{\overline{\rho}}\right)$ disappears, i.e. $\mathbf{R} \to 0$.

**Averaged density evolution equation**

The averaging density evolution equation thank to Fick law gives [10]:

$$\frac{\partial h\overline{\rho}}{\partial t} + \frac{\partial h\overline{u}\overline{\rho}}{\partial x} + \frac{\partial h\overline{v}\overline{\rho}}{\partial y} = \frac{\partial}{\partial x}\left(h\overline{v'\frac{\partial \overline{\rho}}{\partial x}}\right) + \frac{\partial}{\partial y}\left(h\overline{v'\frac{\partial \overline{\rho}}{\partial y}}\right) + \overline{\rho}(\eta) M_1 + \overline{\rho}(Z_b) M_2 + ... \qquad (A9)$$

**Averaged pressure fluctuation equations versus added mass evolution equation**

When the water flow becomes dense, the term $\overline{\rho' \rho'}$ cannot be ignored. For classical non-homogeneous shallow water equations with weakly concentrated flow, this term disappears.

To close the model we write a dynamic equation in terms of $\overline{p'}$ or $\overline{\rho' \rho'}$. Such systems can be obtained by considering the new conservation law equations following:

$$\frac{D\overline{p'}}{Dt} = 0 \qquad (A10)$$

or

$$\frac{D\rho^2}{Dt} = 0 \Rightarrow \frac{D\overline{\rho' \rho'}}{Dt} = 0 \qquad (A11)$$

However, it is important to close by using the conserved variables of the problem.

The equation above describes the evolution of added mass or pressure fluctuations and is a new contribution to the literature.

Based on this conservation layer equation, the two-dimensional equation for added mass evolution can be derived:

$$\frac{\partial h\mathbf{R}}{\partial t} + \frac{\partial h\overline{u}\mathbf{R}}{\partial x} + \frac{\partial h\overline{v}\mathbf{R}}{\partial y} = \frac{\partial}{\partial x}\left(hv'\frac{\partial \mathbf{R}}{\partial x}\right) + \frac{\partial}{\partial y}\left(hv'\frac{\partial \mathbf{R}}{\partial y}\right) + \mathbf{R}(\eta)M_1 + \mathbf{R}(Z_b)M_2 + ... \quad (A12)$$

The final model obtained reads:

$$\begin{cases} \dfrac{\partial h}{\partial t} + \nabla \cdot \left(h\overline{\mathbf{u}}\right) = M_1 + M_2, \\ \dfrac{\partial \overline{\rho h\mathbf{u}}}{\partial t} + \nabla \cdot \left(\overline{\rho}h\overline{\mathbf{u}} \otimes \overline{\mathbf{u}}\right) + \nabla \overline{p} + \nabla \overline{p'} = h\overline{\mathbf{F}} + \rho \mathbf{u}(\eta)M_1 + \rho \mathbf{u}(Z_b)M_2, \\ \dfrac{\partial h\overline{\rho}}{\partial t} + \nabla \cdot \left(h\overline{\mathbf{u}}\overline{\rho}\right) = \nabla \cdot \left(hv'\nabla \overline{\rho}\right) + \overline{\rho}(\eta)M_1 + \overline{\rho}(Z_b)M_2, \\ \dfrac{\partial h\mathbf{R}}{\partial t} + \nabla \cdot \left(h\overline{\mathbf{u}}\mathbf{R}\right) = \nabla \cdot \left(hv'\nabla \mathbf{R}\right) + \mathbf{R}(\eta)M_1 + \mathbf{R}(Z_b)M_2, \end{cases}$$

## Appendix B: Boundary conditions

In an equilibrium state that the rate of settling equals the rate at which the sediment is lifted by turbulent diffusion.

$$c\mathbf{W}_s + \varepsilon_s \frac{\partial c}{\partial z} = 0, \quad z = \eta \quad (B1)$$

The above relation means also that there no sediment exchange at the free surface:

Here, for hyperconcentrated water flow we have $\mathbf{W}_s = (1-c)^n w_s$, with $n = 2.2 + \dfrac{2.65}{1+0.06\mathrm{Res}^{0.7}}$, and with the Reynolds for sediment reads $\mathrm{Res} = \dfrac{w_s d_{50}}{\nu}$. The term $\varepsilon_s$ is the diffusion coefficient obtained from the model and where $c$ is the sediment concentration that influences the mixing density. This term reads:

$$\varepsilon_s = \frac{\varepsilon_f}{S_c}, \quad (B2)$$

where $S_c$ is the Schmidt number that account the phase-lag between sediment turbulent diffusivity of Eddy viscosity. Here this number is assumed non-constant since it depends on settling velocity and bed shear velocity:

$$S_c = \min\left\{1 + 2\left(\frac{w_s}{\mathbf{u}_* + \mathbf{u}_{TOL}}\right), \frac{1}{3}\right\},$$

with $\mathbf{u}_{TOL} = 10^{-6}$ and $\mathbf{u}_* = \dfrac{\kappa \overline{\mathbf{u}}}{\ln(z_0/h) - 1 + (z_0/h)}$, where $z_0 = \dfrac{k_s}{30}$, $k_s$ being the bed roughness coefficient depending on sediment supply condition. To satisfy (B2) we considered two possible alternatives:

(a) $c(x, y, t, z = \eta) = 0$, $\varepsilon_s\big|_{z=\eta} = 0$, (Dirichlet condition)

(b) $\mathbf{W}_s\big|_{z=\eta} = 0$, $\dfrac{\partial c}{\partial z}\bigg|_{z=\eta} = 0$, (Neumann condition)

The sediment flux near the bed is function of sediment entrainment (E) and sediment deposition (D) exchange:

$$c\mathbf{W}_s + \varepsilon_s \frac{\partial c}{\partial z} = D - E, \quad z = Z_b \tag{B3}$$

with

$$\varepsilon_s \frac{\partial c}{\partial z} = -E, \quad z = Z_b \tag{B4}$$